\documentclass[10pt,aps,prb,groupedaddress,showpacs,showkeys]{revtex4-1}

\usepackage{amsmath}
\usepackage{hyperref}
\usepackage{amssymb}
\usepackage{amsfonts}
\usepackage{color}
\usepackage{graphicx}

\usepackage{bm}
\usepackage{float}
\usepackage[justification=justified,singlelinecheck=false]{caption}
\usepackage{subcaption}

\begin{document}
\title{Egocentric physics : just about Mie}
\author{Brian Stout$^{1}$ and Ross McPhedran$^{2}$}
\affiliation{$^1$Aix-Marseille Universit\'e, CNRS, Centrale Marseille, Institut Fresnel, Campus de Saint J\'er\^ome, 13397 Marseille, France }\email[email : ]{brian.stout@fresnel.fr}
\affiliation{$^{2}$ CUDOS, School of Physics, University of Sydney, 2006, Australia.}

\begin{abstract}
We show that the physics of anapole excitations can be accurately described in
terms of a resonant state expansion formulation of standard Mie theory without
recourse to Cartesian coordinate based `toroidal' currents that have
previously been used to describe this phenomenon. In this purely Mie theory
framework, the anapole behavior arises as a result of a Fano-type interference
effect between different quasi-normal modes of the scatterer that effectively
eliminate the scattered field in the associated multipole order.
\end{abstract}

\maketitle
\section{Introduction}
Modern research in metamaterials and nanotechnology often seeks to exploit the
considerable variety in the behaviors of open nano resonator systems. Despite
the complexity of such systems, there has been a growing appreciation of the
fact that their behavior near a resonance can be described in terms of a small
number of parameters within  frameworks
satisfying general physical principles like energy conservation, reciprocity,
and time reversal \cite{dubovik, leung}. There is also an increasing interest  
in exploiting the resonances of high-index dielectrics for metamaterial and 
sensing applications in order to circumvent the problems caused by losses in
metallic resonators at optical and near-optical frequencies \cite{yurinatcomm, wei, yuriopn}. 

An interesting topic arising in the applications and theory of high-index optical 
nanosystems is that of anapole resonances, in which the fields generated by polarization currents inside high index dielectric spheres interfere in such a manner that their superposition 
radiates little energy outside the sphere\cite{yurinatcomm, wei, yuriopn}. Recently, doubt
has been raised as to whether or not the conventional basis of the classic modes of Mie 
scattering theory is well adapted to describe these 
anapole resonances\cite{powell},  or whether it is better
to describe fields in terms of a combination of Mie modes and the alternative 
toroidal modes\cite{dubovik, yurinatcomm,wei}.
Here we will show that the slow convergence of the Mie basis reported by 
Powell \cite{powell} can be readily remedied by incorporating an extra, 
analytically-determined  phase factor, arising in the theory of product 
representations of functions of a 
complex variable\cite{wandw}. With this extra factor included, we show that just 
two resonant modes from Mie theory are sufficient to give an accurate representation 
of the anapole resonance in a case from the literature \cite{yurinatcomm}.

The $\exp\left(-i\omega t\right)$ convention for time harmonic fields will
be used for throughout this work. It is also convenient to express formulas in terms of the 
particle size parameter, $z$,  with respect to the external `background'  media, {\it i.e.} $z\equiv kR$ with 
$k\equiv\frac{\omega}{c} \sqrt{\varepsilon_{r,b}\mu_{r,b}}\equiv N_{b} \frac{\omega}{c}$,
the wavenumber in the background  media.

\section{Anapole `modes'}
\begin{figure}[htb]
\includegraphics[width=0.5\linewidth]{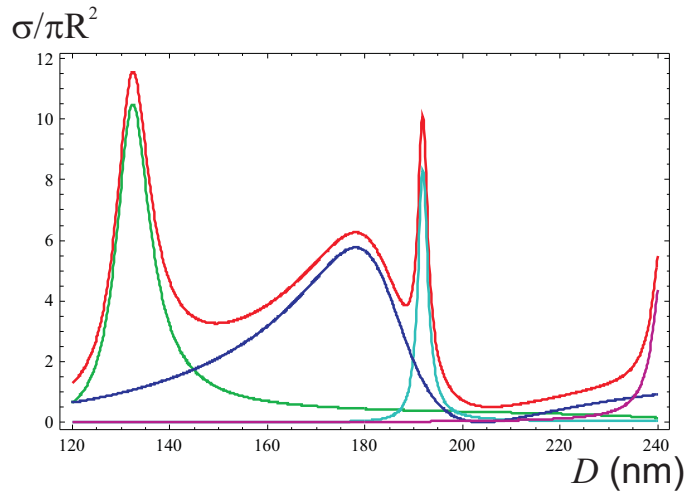}
\caption{Cross section versus particle diameter in nano-meters for a sphere of permittivity,
$\varepsilon_{s}$=16 with a background wavelength of 
$\lambda=550\mathrm{nm}$. 
Total cross section is plotted in red. The electric dipole cross
section exhibiting anapole behavior is plotted in blue, with magnetic dipole contribution
in green and magnetic quadrupole  in cyan. }
\label{Yurcross}
\end{figure}

Anapole `modes' of a scatterer have generally been described in the
context of high index dielectrics. As our first example, we consider a pure
dielectric sphere model with a permittivity index near that of silicon,
$\varepsilon_{s}=16$, for an incident wavelength of $\lambda=550\mathrm{nm}$ that has already 
been shown to possess an anapole behavior\cite{yurinatcomm}. 
The total cross section, $\sigma$, divided by the
geometric cross section, $\pi R^{2}$, plotted in Fig.\ref{Yurcross}, is the sum of all the 
individual multipole contributions of both electric or magnetic wave types. 
The `anapole' condition for the electric dipole mode occurs when 
its contribution to the total cross section is zero, which as can be seen from the figure 
occurs at $D\sim204\mathrm{nm}$.

Each individual contribution to the cross section in Fig.\ref{Yurcross} was calculated from its corresponding $S$-function, $S_{n}$, via the formula,
\begin{equation}
\frac{\sigma_{n}}{\pi R^{2}}=\frac{2n+1}{z^{2}}
\Re \left\{ 1-S_{n}\right\}\;,
\end{equation}
where the $S_{n}$ have exact Mie theory expressions
(given in eq.(\ref{Sdef}) of the appendix\ref{MieTh}). 

The $S_{n}$ functions can alternatively be expressed in terms of a Weierstrass type product expansion
in terms of the resonant state frequencies, $\omega_{\alpha}$, or equivalently resonant state size parameters, $z_{\alpha}\equiv k_{\alpha}R\equiv N_{b}\omega_{\alpha}R/c$.
For each multipolar order and wave type, the $z_{\alpha}$ can be readily 
determined as zeros of the transcendental equations in eq.(\ref{den_funcs}), with the $S_n$
function taking the form,
\begin{align}
S_{n}\left(  z\right) &=(-)^{n+1}e^{-2iz}\prod\limits_{\alpha=-\infty}^{\infty} 
\frac{z-z_{n,\alpha}^{\ast}}{z-z_{n,\alpha}} \label{Sprodz}\\
 &= (-)^{n+1}e^{-2iz}\prod\limits_{\alpha=1}^{\infty}
 \frac{z^{2}-\left\vert z_{n,\alpha}\right\vert^{2}
 +2iz{\rm Im}\left(z_{n,\alpha}\right)}{z^{2}-\left\vert z_{n,\alpha}\right\vert^{2}
 -2iz{\rm Im}\left(z_{n,\alpha}\right)}\;, \nonumber
\end{align}
where the $e^{-2iz}$  phase factor is associated with causality. (This phase factor
can be expressed either in terms of a sum over the difference between the $z$ values 
of zeros and poles, summed over the entire set, or obtained in a more general way using causality - 
see King \cite{king} Vol.2.)
In the first line of eq.(\ref{Sprodz}), time-reversal symmetry requires a
symmetry between negative and positive resonant state eigenvalues,
 \begin{equation}
-z_{n,-\alpha}^{\ast}=z_{n,\alpha}\;,
\end{equation}
while in the last line of eq.(\ref{Sprodz}) this symmetry is exploited to remove 
explicit reference to the negative energy resonant states.
We remark that eq.(\ref{Sprodz}) is only valid as it stands for
dielectric scatterers (the zeros of $S_{n}$ are not
the complex conjugates of the resonant state values for dispersive media), and that
eq.(\ref{Sprodz}) is the form applicable to electric (TM) modes 
(\emph{magnetic (TE)} product expressions are in fact identical 
\emph{but of opposite sign} with a $(-)^{n}$ prefactor).
\begin{figure}[htb]
\centering\includegraphics[width=0.5\linewidth]{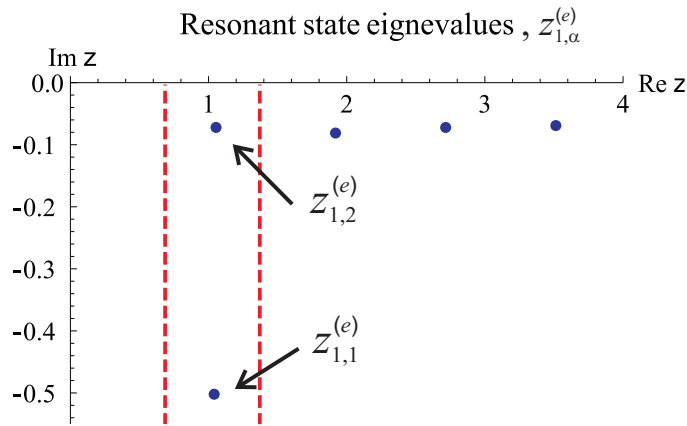}\caption{Size parameter
eigenvalues of the electric dipole resonant states, $z_{1,\alpha}^{(e)}$, for
a sphere of permittivity, $\varepsilon_{s}$=16. Vertical dashed lines delimit
the size parameter range evaluated in Fig.\ref{Yurcross}}
\label{QNM_product}
\end{figure}
\begin{figure}[htb]
\centering\includegraphics[width=0.5\linewidth]{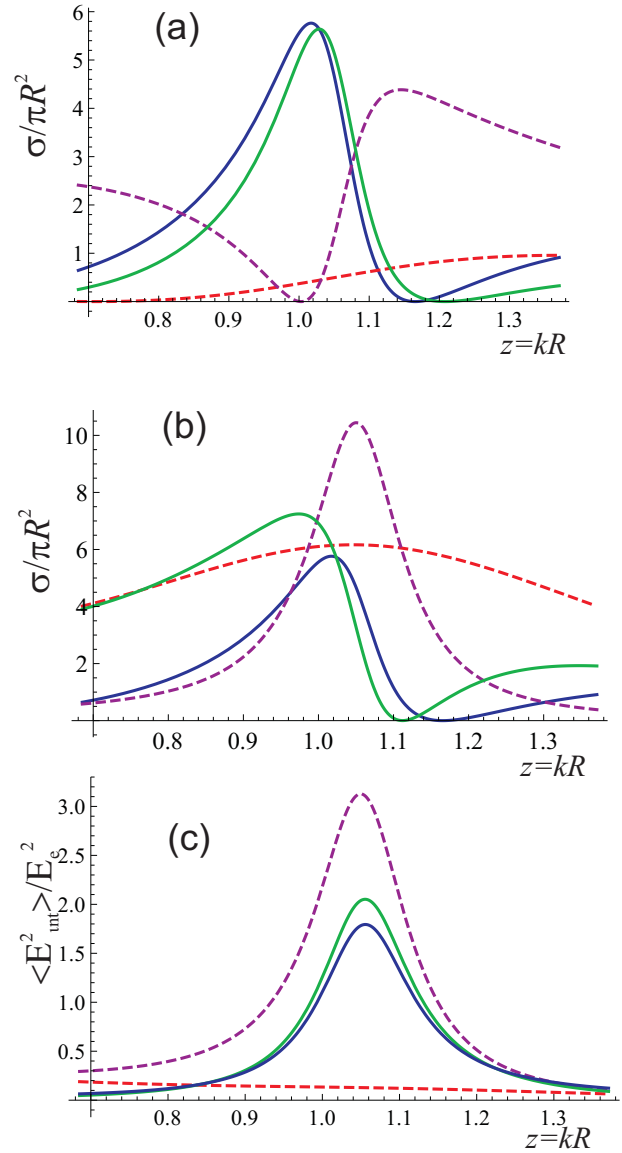}\caption{Electric dipole 
cross section, (a)-(b), and average internal field enhancement, (c). Total contributions as blue curves. 
Dashed red curve in (a) and (c) corresponds to including only the 
$z_{1,1}^{(e)}$ resonant state, while in (b) it corresponds to a superposition of the non-resonant
contribution and the $z_{1,1}^{(e)}$ resonant state. 
The dashed purple lines in (a)-(c) correspond to inclusion of only the $z_{1,2}^{(e)}$ contributions. 
The green curves correspond to a coherent sum of $z_{1,1}^{(e)}$ and $z_{1,2}^{(e)}$ resonant states.}
\label{QNM_decomp}
\end{figure}

From here on we concentrate attention on the electric dipole response since it
exhibits anapole behavior. The electric dipole resonant states with the six lowest 
real parts are plotted in Fig.\ref{QNM_product} where the $z_{n,\alpha}$ are
obtained by solving for the zeros of the transcendental expression in
eq.(\ref{dene_func}). We immediately see from Fig.\ref{QNM_product} that
there are two electric dipole resonant states in the diameter range explored
in Fig.\ref{Yurcross} (region between the dashed vertical lines). Their values
are $z_{1}\simeq1.039-i0.501$, with a rather large imaginary part and
$z_{2}\simeq1.053-i0.0723$ close to the real $z$ axis. 

The total electric dipole cross section (blue curve) is plotted in Fig.\ref{QNM_decomp},
over the same $D=120$ to 240nm range at $\lambda_{0}=550\mathrm{nm}$, as in Fig.\ref{Yurcross}
but with size parameter indicated on the ordinate in order to facilitate the comparison
with Fig.\ref{QNM_product}. The blue curve is the total electric dipole cross
section. The red curve in Fig.\ref{QNM_decomp} is obtained if we only include
the $z_{1}$ contribution to the product expansion of $S_{1}^{\left(  e\right)}\left(z\right)$, while 
the green curve is obtained by including both $z_{1}$ and $z_{2}$ in the $S$-function product 
expansion of eq.(\ref{Sprodz}).
The addition of the higher resonant states makes $\sigma_{1}^{(e)}\left( z\right)$ 
rather slowly converge to the exact blue curve. 

\begin{figure}[tbh]
\centering\includegraphics[width=0.5\linewidth]{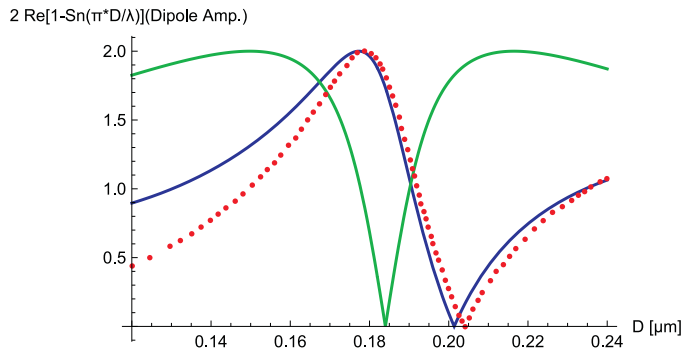}\caption{
The product form expression based on the two dipole mode expression (\ref{twoterm}) (blue curve)
and the same without the exponential factor (green curve), compared with  digitized points from  Fig.2 of
Miroschnichenko   \emph{et al} \cite{yurinatcomm} (red points). 
Diameter varies from 120 to 240 nm at $\lambda_{0}=550\mathrm{nm}$.}
\label{cfnatcomm}
\end{figure}

We now compare the expression (\ref{Sprodz}) restricted to just the two dipole terms of the previous paragraph,
\begin{equation}
\sigma_1(z)\propto 2 \Re \left[1- \exp (-2 i z ) \frac{(z - z_{1,1}^{\ast}) (z - z_{1,2}^{\ast})}{(z - z_{1,1}) (z - z_{1,2})} \right] \;, 
\label{twoterm}
\end{equation} 
with the calculations of
Miroschnichenko A.E.  \emph{et al} \cite{yurinatcomm} incorporating both Cartesian electric and toroidal  dipoles- see Fig. \ref{cfnatcomm}.
The simple expression (\ref{twoterm}) with the exponential factor present does very well in predicting the whole of the curve
after the resonant maximum, through the anapole minimum and right to the large diameter end of the figure. Without the exponential factor, the two-term theory is totally inadequate. With the exponential factor, there is no need to incorporate toroidal terms
in the Mie theory, in predicting the anapole minimum.

Although Figs.\ref{QNM_decomp}(a) and \ref{cfnatcomm} establish that both the $z_{1}$ and 
$z_{2}$ resonant states are required to reproduce the anapole behavior, product expansions don't 
readily lend themselves to the more familiar explanations of the Fano shape as an interference effect.
This more traditional reasoning is obtainable by reformulating the cross section 
contributions as a resonant state sums of the Mie `coefficients', $a_{n}$ and $b_{n}$, 
(which are opposite in sign from the $T_{n}$-functions), given in eq.(\ref{Sdef}) of the appendix. 
The $T_{n}$-functions by definition relate the coefficients of the incident field to those of the scattered field as in eq.(\ref{Trel}), and are
consequently related to the $S_{n}$ functions via $S_{n}=1+2T_{n}$, and 
contribute to the scattering cross section via the relation,
\begin{equation}
\frac{\sigma_{n}}{\pi R^{2}}=\frac{2n+1}{z^{2}}
2\left\vert T_{n}\right\vert ^{2}\;.
\end{equation}
The $T_{n}$ functions can be expressed as a the sum of a non-resonant term 
plus a sum over the resonant state responses,
\begin{align}
T_{n}(z) &=\frac{(-)^{n+1}e^{-2iz}-1}{2} +\frac{ (-)^{n+1}e^{-2iz}}{2}
\sum_{\alpha=-\infty}^{\infty}\frac{r_{n,\alpha}}{z-z_{n,\alpha}}
\nonumber\\
&= \frac{(-)^{n+1}e^{-2iz}-1}{2} 
\label{TLor} \\ 
& +(-)^{n+1}e^{-2iz}\sum_{\alpha=1}^{\infty}\frac{iz{\rm Im}
\left( r_{n,\alpha}\right) +{\rm Re}\left( r_{n,\alpha}z_{n,\alpha}^{\ast}\right)}
{z^{2}-\left\vert z_{n,\alpha}\right\vert^{2}-2iz{\rm Im}\left(z_{n,\alpha}\right)  }\;, \nonumber
\end{align}
where in the last line we used the fact that time reversal symmetry imposes the following
relation on the residue factors,
\begin{equation}
r_{n,-\alpha}=-r_{n,\alpha}^{\ast}\;,
\end{equation}
in order to eliminate summing over the negative energy states.
 
We remark that unlike the product form of eq.(\ref{Sprodz}),
the sum formulation of eq.(\ref{TLor}) requires a calculation of the residues, $r_{n,\alpha}$.
The $r_{n,\alpha}$ can be obtained in a number of different manners:
directly from eq.(\ref{Sprodz}) if one has determined
a sufficiently high number of resonant states, analytically via Mie theory and the 
Cauchy principal value theorem as given in eq.(\ref{Tresid}), or from far field information.
For the same, $\varepsilon_{s}=16$, permittivity as in Figs.\ref{Sprodz} and
\ref{QNM_decomp}, the residues associated with the first two resonant states
are $r_{1}=-0.474-1.740$ and $r_{2}=0.0376+0.1996$ so that even though the
$z_{1}$ mode is far from the real axis, it cannot be neglected on account of
its large residue strength. We plot the total electric dipole cross section in
Fig.\ref{QNM_decomp}. The red curve is obtained by taking the sum in
eq.(\ref{TLor}) of the non-resonant contribution plus the broad $z_{1}$
resonant state contribution. Adding in the the $z_{2}$ resonant state 
contribution results in the green curve while the purple curve was obtained by treating the
$z_{2}$ contribution in isolation. We thus see that the anapole behavior has been
predominantly recovered as a Fano type interference phenomenon between narrow
and broad modes. To fully understand anapoles, we need to 
see how the field inside the particle is modified with respect to that which would
exist in its absence, as discussed below.

\section{Internal field averages}

The functions determining the linear relationship between internal and excitation field coefficients, 
are denoted $\Omega_{n}$, cf. eq.(\ref{Omegdef}) 
(these are commonly called the $c_{n}$ and $d_{n}$ coefficients in Mie theory for TM 
and TE modes respectively). 
They can be written as a sum
of resonant state Lorentzians like the $T$-matrix expression in eq.(\ref{TLor}), 
(without the non-resonant contributions),
\begin{align}
\Omega_{n}&=\frac{(-)^{n+1}e^{-2iz}}{2}\sum_{\alpha=-\infty}^{\infty}
\frac{r_{n,\alpha}^{\left(\Omega\right)}}{z-z_{n,\alpha}} \nonumber \\
& \!\!\!\!\!\!\!\!\!\! =  \left(-\right)^{n+1}e^{-2iz}\sum_{\alpha=1}^{\infty}\frac{iz{\rm Im}
\left(r_{n,\alpha}^{\left(\Omega\right)}\right)  
+{\rm Re}\left(r_{n,\alpha}^{\left(\Omega\right)}z_{n,\alpha}^{\ast}\right)}
{z^{2}-\left\vert z_{n,\alpha}\right\vert^{2}-2iz{\rm Im}
\left(z_{n,\alpha}\right)} \;. \label{OmegLor}
\end{align}
The residue factors, $r_{n,\alpha}^{\left(\Omega\right)}$, differ from those of 
the $T$-matrix, and can be calculated analytically in Mie theory, and are given in eq.(\ref{Intresid}). 

Although we know of no precise analogue of cross section for the internal field,
the volume average of the internal field intensity divided by excitation field intensity
at the spheres center is a physically relevant quantity which 
also has additive multipole contributions,
\begin{align}
\frac{\left< || \boldsymbol{E}^{\rm (int)} ||^{2} \right>_{V_{s}}}{ || \boldsymbol{E}_{\rm e}(\boldsymbol{0}) ||^{2}} = & \nonumber \\
& \!\!\!\!\!\!\!\!\!\!\!\!\!\!\!\!\!\!\!\!\! \frac{3}{2 z^{3}}\sum_{n=1}^{\infty} \left(2n+1\right) \left\{I_{n} |\Omega_{n}^{(e)}|^{2} 
+ J_{n}|\Omega_{n}^{(h)}|^{2} \right\}  \;,
\end{align}
where $I_{n}$ and $J_{n}$ arise from analytical integrations of the field squared, 
(see eq.(\ref{I_J_funcs}) of the appendix). We see in Fig.(\ref{QNM_decomp}c) that the isolated
contribution of the $z_{1,1}^{(e)}$ resonant state to the internal field intensity is considerably smaller
than that of the $z_{1,2}^{(e)}$ state, but their superposition is still required to reasonably
approximate the total average field intensity.
We remark from Fig.(\ref{QNM_decomp}c) that the internal field resonance is still significant 
at the anapole size parameter of $z_{\rm anap}\simeq 1.16$.

\section{Resonant state field expansions\label{Field}}

As we have seen above, physically relevant quantities like cross sections 
and field averages can be formulated entirely in terms of the resonant frequencies and residues
without explicit reference to the resonant state fields themselves. 
Nevertheless, in applications, like LDOS, Lamb shift, and the present
analysis of ``toroidal'' currents, it is desirable to 
develop the field distribution directly in terms of the resonant states.

Mie theory provides an analytic expression for the multipolar component of the electric field inside the 
sphere originating from electric modes, $\boldsymbol{E}^{({\rm int},e)}_{n,m}$, which  can be determined from eqs.(\ref{Eint}) and (\ref{Omegdef}),
\begin{equation}
\boldsymbol{E}^{({\rm int},e)}_{n,m} = E e_{n,m} \Omega_{n}^{(e)} k^{3/2}\boldsymbol{N}_{n,m}^{\left(1\right)}\left( k_{s}\boldsymbol{r}\right)  \;, \label{EMie}
\end{equation}
where $e_{n,m}$ is the multipolar excitation field coefficient determined
for a field of amplitude of $k^{3/2}$ at the origin, and $E$ is a field
amplitude parameter\cite{stout08}.
The resonant state expansion of $\boldsymbol{E}^{\rm(int)}_{n,m}$ is,
\begin{equation}
\boldsymbol{E}_{n,m}^{({\rm int},e)} = \frac{(-)^{n+1}e^{-2iz}Ee_{n,m}}{2}
\sum_{\alpha=-\infty}^{\infty}
\frac{r_{n,\alpha}^{\left(\Omega\right)}}{z-z_{n,\alpha}}
\boldsymbol{E}_{n,m,\alpha}^{({\rm int},e)}\label{Ealph}\;, 
\end{equation}
where $\boldsymbol{E}_{n,m,\alpha}^{({\rm int},e)}$ 
are the resonant state wave functions of the electric $n,m$ multipolar mode, 
for which the analytical expressions in the case of a homogeneous sphere are,  
\begin{align}
  \boldsymbol{E}_{n,m,\alpha}^{({\rm int},e)}(\boldsymbol{r}) &=
  \left[\sqrt{n\left(  n+1\right)}j_{n}
  \left(\rho_{s} z_{n,\alpha}\tilde{r} \right)  
  \boldsymbol{Y}_{n,m}(\theta,\phi) \right. \nonumber\\
  & \left. \!\!\!\!\!\!\!\!\!\!\!\!\!\! +\psi_{n}^{\prime}\left( \rho_{s}z_{n,\alpha} \tilde{r}\right)  
  \boldsymbol{Z}_{n,m}(\theta,\phi)\vphantom{\sqrt{n\left(n+1\right)}}\right]
  \frac{ k_{n,\alpha}^{3/2} }{\rho_{s} z_{n,\alpha}\tilde{r} }  \;, \label{resdev}
\end{align}
where $\boldsymbol{Y}_{n,m}$ and $\boldsymbol{Z}_{n,m}$ are normalized 
vector spherical harmonics\cite{stout05}.
In eq.(\ref{resdev}), $\rho_{s}$ is the 
refractive index contrast of the sphere (cf. eq.(\ref{rhos})), and
$\tilde{r}\equiv r/R$ is a dimensionless radial coordinate normalized with respect 
to the sphere's radius ($\tilde{r}\le 1$).
\begin{figure}[htb]
\centering\includegraphics[width=0.5\linewidth]{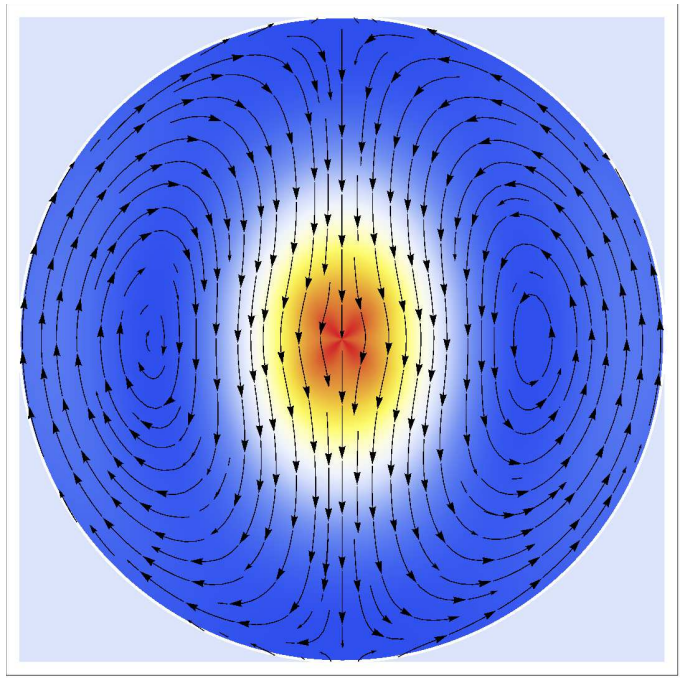}\caption{Real part of the electric field lines (arrows) from Mie theory 
(\emph{i.e.} from eqs.(\ref{Ealph}) or (\ref{resdev})) 
plotted on a background (colour- red maximum) of the internal electric field intensity plot excited
by a $n=1$, $m=0$ electric incident field at the anapole point of $D=204$nm for a wavelength of
$\lambda_{b}=550\mathrm{nm}$.}
\label{fldlines}
\end{figure}

The real part of the electric field lines from Mie theory are plotted in 
Fig.\ref{fldlines} on a background of an electric field intensity plot at the anapole point. 
One sees that there is indeed a ``toroidal'' aspect to the induced polarization currents, but these
are completely described  within the context of Mie theory without the need for any additional
basis functions.

\section{Conclusions\label{Concl}}
The approach we have described to achieve an understanding of the scattering properties of high index dielectric spheres, and
in particular of their anapole excitations, has a number of advantages. Firstly, it relies only on the knowledge of the positions of the
relevant complex resonances calculated from Mie theory. Secondly, it incorporates an analytically determined phase term, without which a very large number of terms in the product over resonances would be required to achieve accuracy. Thirdly, it does not require the evaluation of inner products or normalisation factors for different resonant modes, which have proved problematic for other approaches. We have shown that, with a small number of the terms relevant to a given sphere diameter range, it can yield information about scattering behaviour of good accuracy, with the simplicity of the formulae used being beneficial to the physical understanding of
the phenomena under investigation. The approach is quite general, and will no doubt find applications in a wide range of situations
in addition to those studied here.

\acknowledgments
Research conducted within the context of the International Associated Laboratory for Photonics
between France and Australia. This work has been carried out thanks to the support of the A*MIDEX
project (no. ANR-11-IDEX-0001-02) funded by the Investissements d'Avenir French Government program, managed by the French National Research Agency (ANR). The authors would like to than
Remi Colom, Emmanuel Lassalle, and Nicolas Bonod for helpful discussions.

\section{Appendix: Mie theory\label{MieTh}}

The excitation, $\boldsymbol{E}_{\mathrm{e}}$, and scattered, 
$\boldsymbol{E}_{\mathrm{s}}$, fields can respectively be developed as a multipolar decomposition,
\begin{align}
\boldsymbol{E}_{\mathrm{e}}\left(  k\boldsymbol{r}\right)   &  
=E\sum_{n,m} \left[  e_{nm}^{\left(  e\right)}
\boldsymbol{N}_{nm}^{\left(  1\right)  }(k\boldsymbol{r})
+e_{nm}^{\left(h\right)}
\boldsymbol{M}_{nm}^{\left(  1\right)  }(k\boldsymbol{r})\right]
\nonumber\\
\boldsymbol{E}_{\mathrm{s}}\left(  k\boldsymbol{r}\right)   &  =
E\sum_{n,m}\left[  f_{nm}^{\left(  e\right)}
\boldsymbol{N}_{nm}^{\left(+\right)  }(k\boldsymbol{r})
+f_{nm}^{\left(h\right)  }\boldsymbol{M}_{nm}^{\left(+\right)}(k\boldsymbol{r})\right]
\;, \label{Eexp}
\end{align}
where $n$ is the electromagnetic angular quantum number, and 
$m$ the projection quantum number. The real parameter,
$E$, determines the strength of the incident field, while multipole wave
functions, $\boldsymbol{N}_{n,m}(k\boldsymbol{r})$ and 
$\boldsymbol{M}_{n,m}(k\boldsymbol{r})$, are respectively of the `electric' and 
`magnetic' types (often called TM and TE type fields)\cite{stout05}, and their
coefficients are respectively distinguished with $\left(e\right)$ and 
$\left(h\right)$ superscripts. The superscript, $^{(1)}$ on the multipolar 
fields in eq.(\ref{Eexp}) denotes regular type fields described by spherical Bessel
functions, whereas $^{(+)}$, $^{(-)}$ refers to fields with outgoing (incoming)
boundary conditions (Hankel type fields).

The $T_{n}$ functions of Mie theory are meromorphic functions 
of frequency and constitutive parameters that relate the scattered
field coefficients, $f_{nm}$, to excitation field coefficients, $e_{nm}$,
\begin{equation}
f_{nm}=T_{n} e_{nm} \;. \label{Trel}
\end{equation}

Although incident and scattered fields are very useful from a response function 
standpoint, it is also to work with the \emph{total fields} that are actually present in the
particle plus field system. The partial wave representations of the total field 
inside the particle, $\boldsymbol{E}_{\mathrm{int}}$, and the total external,
field, $\boldsymbol{E}_{\mathrm{t}}$, are respectively,
\begin{subequations}
\begin{align}
\boldsymbol{E}_{\mathrm{int}}   &
=E\sum_{n,m} \left[  s_{nm}^{\left(  e\right)}
\boldsymbol{N}_{nm}^{\left(1\right)}(k_{s}\boldsymbol{r})
+s_{nm}^{\left(  h\right)  }\boldsymbol{M}_{nm}^{\left(1\right)}
(k_{s}\boldsymbol{r})\right]  \label{Eint}\\
\boldsymbol{E}_{\mathrm{t}}  &  =
E\sum_{n,m}\left[a_{+,nm}^{\left(e\right)}
\boldsymbol{N}_{nm}^{\left(+\right)}(k\boldsymbol{r})
+a_{+,nm}^{\left(h\right) }\boldsymbol{M}_{nm}^{\left(+\right)}
(k\boldsymbol{r})\right. \nonumber\\
& \qquad +\left.  a_{-,nm}^{\left(e\right)}\boldsymbol{N}_{nm}^{\left(-\right)}
(k\boldsymbol{r})+a_{-,nm}^{\left(h\right)}
\boldsymbol{M}_{nm}^{\left(-\right)}(k\boldsymbol{r})\right]  \;.
\end{align}
\end{subequations}

The linear relationship between the internal field coefficients
and the incident field coefficients is denoted by the matrix, $\Omega$, which is
diagonal for spherically symmetric problems, For spheres, the internal field response functions, 
$\Omega_{n}$, relate the internal field coefficients, $s_{n,m}$, to the excitation field coefficients, $e_{n,m}$,
\begin{equation}
s_{n,m}=\Omega_{n} e_{n,m}\;, \label{Omegdef}
\end{equation}
while the $S$-matrix relates incoming to outgoing components of the 
total external field,
\begin{equation}
a_{+,n,m}=
S_{n} a_{-,n,m} \;.
\end{equation}
The $T$, $S$, and $\Omega$ functions have exact expressions in
Mie theory, 
\begin{equation}
\Omega_{n} =\frac{1}{D_{n}}, \quad S_{n} =-\frac{N_{S,n}}{D_{n}},
 \quad T_{n} =-\frac{N_{T,n}}{D_{n}} \;, \label{Sdef}
\end{equation}
where the numerator functions can be expressed,
\begin{subequations}
\begin{align}
N_{S,n}^{\left(  e\right)}   &  \equiv z
\frac{\varepsilon_{s}j_{n}\left(\rho_{s}z\right) \xi_{-,n}^{\prime}\left(z\right)
-\psi_{n}^{\prime}\left(\rho_{s}z\right)  h_{-,n}\left(z\right)}
{i\rho_{s}}\\
N_{S,n}^{\left(h\right)}   &  \equiv z\frac{\mu_{s} j_{n}\left(\rho_{s}z\right)  
\xi_{-,n}^{\prime}\left(z\right)  -\psi_{n}^{\prime}\left(\rho_{s}z\right)  
h_{-,n}\left(z\right)}{i\mu_{s}}\;, \label{numS_func} \\
N_{T,n}^{\left(  e\right)  }   &  \equiv z\frac{\varepsilon
_{s}j_{n}\left(  \rho_{s}z\right)  \psi_{n}^{\prime}\left(  z\right)
-\psi_{n}^{\prime}\left(  \rho_{s}z\right)  j_{n}\left(  z\right)}
{i\rho_{s}}\\
N_{T,n}^{\left(  h\right)  }  &  \equiv z\frac{\mu_{s}%
j_{n}\left(  \rho_{s}z\right)  \psi_{n}^{\prime}\left(  z\right)  -\psi
_{n}^{\prime}\left(  \rho_{s}z\right)  j_{n}\left(  z\right)  }{i\mu_{s}}\;,
\label{numT_func}%
\end{align}
\end{subequations}
and `denominator' functions expressed, 
\begin{subequations}
\begin{align}
D_{n}^{\left(  e\right)}   &  \equiv z\frac{\varepsilon_{s}
j_{n}\left(  \rho_{s}z\right)  \xi_{+,n}^{\prime}\left(  z\right)
-\psi_{n}^{\prime}\left(  \rho_{s}z\right)  h_{+,n}\left(  z\right)}
{i\rho_{s}}\label{dene_func}\\
D_{n}^{\left(  h\right)}   &  \equiv z\frac{\mu_{s}j_{n}
\left(\rho_{s}z\right) \xi_{+,n}^{\prime}\left(z\right) 
 -\psi_{n}^{\prime}\left( \rho_{s}z\right)  h_{+,n}\left( z\right)}{i\mu_{s}}\;.
\label{denh_func}
\end{align}
\label{den_funcs}
\end{subequations}
In Mie theory, the internal field coefficient factors
are traditionally defined, $\Omega_{n}^{\left(e\right)}\equiv d_{n}$
and $\Omega_{n}^{\left(h\right)}\equiv c_{n}$.

The $\psi_{n}$ and $\xi_{n,\pm}$ are Riccati spherical Bessel functions, 
$\psi_{n}\left(z\right)\equiv zj_{n}\left(z\right)$, and 
$\xi_{n,\pm}\left(  z\right)\equiv zh_{n}^{\left(\pm\right)}\left(z\right)$.
Here, the $\varepsilon_{s}$ and $\mu_{s}$ are the relative constitutive parameters with
respect to the background media,
\begin{equation}
\mu_{s}\equiv\frac{\mu_{r,s}}{\mu_{r,b}}\qquad\varepsilon_{s}\equiv
\frac{\varepsilon_{r,s}}{\varepsilon_{r,b}}\;,
\end{equation}
and $\rho_{s}$ is the refractive contrast with respect to the background
medium,
\begin{equation}
\rho_{s}\equiv\sqrt{\varepsilon_{s}\mu_{s}}=
\sqrt{\frac{\varepsilon_{r,s}\mu_{r,s}}{\varepsilon_{r,b}\mu_{r,b}}} \label{rhos} \;.
\end{equation}

Given the Cauchy residue theorem, a comparison of eqs.(\ref{OmegLor})
and (\ref{Sdef}), leads to the following
residues of the internal field coefficients, 
$r_{n,\alpha}^{\left(\Omega,e,h\right)}$,
\begin{equation}
r_{n,\alpha}^{\left(\Omega\right)}=\frac{2e^{2iz_{n,\alpha}}}
{\left.  \frac{d}{dz}D_{n}\left(  z\right)  \right\vert_{z=z_{n,\alpha}}}\;,
\end{equation}
which following analytic manipulations results in,
\begin{align}
\frac{2i\rho_{s}e^{2iz_{n,\alpha}}}{r_{n,\alpha}^{\left( \Omega,e\right)}}  & =
z_{\alpha}^{2}h_{+,n}\left(  z_{\alpha}\right) j_{n}
\left(\rho_{s}z_{\alpha}\right) \varepsilon_{s} \left(\mu_{s}-1\right)
\nonumber \\ &\!\!\!\!\!\!\!\!\!\!\!\!\!\!\!\!\!\!\!\!\!\!\!+\left( {\varepsilon_{s}-1} \right) \left[
\xi_{+,n}^{\prime}\left(  z_{\alpha}\right)  \psi_{n}^{\prime}\left(  \rho_{s}z_{\alpha}\right)  
\right. \nonumber \\  & \left. \vphantom{ \xi_{+,n}^{\prime}}  \!\!\!\!\!\!\!  +n\left( n+1\right) h_{+,n}\left(  z_{\alpha}\right)  j_{n}\left(\rho_{s}
 z_{\alpha}\right)  \right]  
 \nonumber\\
\frac{2i \mu_{s}e^{2iz_{n,\alpha}}}{r_{n,\alpha}^{\left( \Omega,h\right)}}  &
=z_{\alpha}^{2}h_{+,n}\left(  z_{\alpha}\right)  j_{n}
\left(\rho_{s}z_{\alpha}\right) \mu_{s} \left(  \varepsilon_{s}-1\right) 
\nonumber \\ &\!\!\!\!\!\!\!\!\!\!\!\!\!\!\!\!\!\!\!\!\!\!\!\!\!\!\!\!
\quad+\left( \mu_{s}-1 \right) \left[  \xi_{+,n}^{\prime}\left( z_{\alpha}\right)  \psi_{n}^{\prime}
\left(\rho_{s}z_{\alpha}\right) \right. \nonumber \\  & \!\!\!\!\!\!\! \left. \vphantom{ \xi_{+,n}^{\prime}}+h_{+,n}\left(  z_{\alpha}\right)  j_{n}
\left(\rho_{s}z_{\alpha}\right)  n\left( n+1\right)  \right]   \;, \label{Intresid}
\end{align}
for electric and magnetic multipole resonant states respectively.  
Analogous comparisons give for the residues, $r_{n,\alpha}$, of the $T$-matrix in eq.(\ref{TLor}), 
\begin{align}
r_{n,\alpha} &  =-\frac{2 N_{T,n}\left(  z_{n,\alpha}\right)ie^{2iz_{n,\alpha}}  }
{\left. \frac{d}{dz}D_{n}\left(  z\right)\right\vert _{z=z_{n,\alpha}}}\equiv \frac{2ie^{2iz_{n,\alpha}}}
{\mathcal{N}^{2}_{n,\alpha }}\; , \label{Tresid}
\end{align}
where $\mathcal{N}_{n,\alpha}$ correspond to the resonant state 
``normalisation\textquotedblright\ factors and for the electric and magnetic modes  
are respectively given by,
\begin{align}
\left[  \mathcal{N}_{n,\alpha}^{\left(  e\right)}\right]^{2}  &=
\xi_{+,n}^{2}\left(  z_{n,\alpha}^{\left( e\right)}\right)
\left(  \mu_{s}-1\right)  \nonumber\\
 & \!\!\!\!\!\!\!\!\!\!\!\!\!\!\!\!\!  +\left\{[\xi_{+,n}^{\prime}\left(  z_{n,\alpha}^{\left( e\right)}\right)]^{2} 
+\frac{n\left(  n+1\right)}{\varepsilon_{s}}h_{+,n}^{2}
\left(z_{n,\alpha}^{\left(  e\right)}\right)  \right\}  
\left( \varepsilon_{s}-1\right)  \nonumber \\ 
\left[ \mathcal{N}_{n,\alpha}^{\left(  h\right)}\right]^{2}  & 
=  \xi_{+,n}^{2}\left( z_{n,\alpha}^{\left( h\right)}\right)  
\left(\varepsilon_{s}-1\right) \label{Norms}\\
 &  \!\!\!\!\!\!\!\!\!\!\!\!\!\!\!\!\! +\left\{\left[ \xi_{+,n}^{\prime}\left(  z_{\alpha}^{\left( h\right)}\right)\right]^{2}
+\frac{n\left(  n+1\right)}{\mu_{s}}h_{+,n}^{2}\left(z_{n,\alpha}^{\left( h\right)}\right)\right\}
  \left(\mu_{s}-1\right) \;.\nonumber
\end{align}
which given the differences in notation and conventions 
simplify to the normalization expressions derived in ref.\cite{Mulj16} when restricted to the
special case of null permeability contrast, {\it i.e.} $\mu_{s}=1$.

It is important to keep in mind that resonant states have important 
and somewhat unfamiliar properties when compared with the much more 
common Hermitian spectral analysis of closed conserving systems. Notably,
residue/normalisation factors like $\mathcal{N}_{n,\alpha}$, are complex 
numbers so that 
$\mathcal{N}_{n,\alpha}^2 \neq |\mathcal{N}_{n,\alpha}|^{2}$). 

The expressions $I_{n}$ and $J_{n}$ for determining the average internal field in eq.(\ref{QNM_decomp}) 
are found by analytically evaluating the following integrals for the internal fields:
\begin{subequations}
\begin{align}
\frac{2\rho_{s}^{2}}{z} I_{n} & \equiv  \frac{2\rho_{s}^{2}}{z}  \int_{0}^{z}
\left\{  j_{n}^{2}\left(  \rho_{s}\eta\right) n
 \left(  n+1\right)  +\psi_{n}^{\prime,2}\left(  \rho_{s}\eta\right) \right\} d\eta \nonumber \\ 
& \!\!\!\!\!\!\!\!\!\!\!\!\!\!\!\!\! = \psi_{n}^{\prime}\left(  \rho_{s}z\right)
 \left[ \psi_{n}^{\prime}( \rho_{s}z\right)+j_{n}\left(  \rho_{s}z )\right] \nonumber\\
 & \qquad +  (\rho_{s}^2z^2 -n\left(  n+1\right)) j_{n}^{2}\left(  \rho_{s}z\right) \\
\frac{2\rho_{s}^{2}}{z} J_{n}  & \equiv  \frac{2\rho_{s}^{2}}{z} 
\int_{0}^{z} j_{n}^{2}\left(  \rho_{s}\eta\right)\eta^{2}d\eta \nonumber\\
 &\!\!\!\!\!\!\!\!\!\!\!\!\!\!\!\!\! = \psi_{n}^{\prime}\left(\rho_{s}z\right)
  \left[ \psi_{n}^{\prime}( \rho_{s}z\right)-j_{n}\left(  \rho_{s}z )\right]
   \nonumber\\
 & \qquad +  (\rho_{s}^2z^2 -n\left(  n+1\right)) j_{n}^{2}\left(  \rho_{s}z\right)  \;. \label{I_J_funcs}
\end{align}
\end{subequations}

%\begin{figure}[H]
%\includegraphics[width=1\columnwidth]{Optimal_Fig1.png}
%\caption{ Approximate values for the electric dipole Mie coefficient, $a_{1}$, (amplitude and phase - dashed blue), compared with the exact values (solid black) ; $R=60$-nm gold %sphere.}
%\label{a1_gold_sphere}
%\end{figure}

\end{document}